\documentclass[pra,twocolumn,showpacs,showkeys,preprintnumbers,amsmath,amssymb]{revtex4-2}
\makeatletter
\usepackage[dvips]{graphicx}	

\usepackage{amsmath}
\usepackage{amssymb}
\usepackage{amsbsy}
\usepackage{color}
\setcitestyle{square}
\usepackage{epstopdf}
\DeclareGraphicsExtensions{.pdf,.eps,.png,.jpg,.mps}

\begin{document}

\title{
Topological electric field-defined quantum dots in bilayer graphene: 
\\An atomistic approach
}
\author{ W. Jask\'olski} 
\email{wj@fizyka.umk.pl}
\affiliation{Institute of Physics, Faculty of Physics, Astronomy, and Informatics, Nicolaus Copernicus University in Torun, Grudziadzka 5, 87-100 Toru\'n, Poland}

\begin{abstract}
We study topological bound states in quantum dots defined by an electric field in bilayer graphene. 
An external field is perpendicular to the bilayer, and changes sign in a finite region that defines the quantum dot. The electric field opens a gap in the bilayer graphene 
and the reversed field creates a domain wall with one-dimensional chiral gapless bands localized therein. 
The finite size of dots leads to the quantization of these bands and the appearance of discrete bound states 
localized at the dot boundary. We consider rectangular dots oriented along the armchair and zigzag directions. 
We go beyond a simple continuum one-valley model and use an atomistic tight-binding approach. 
This allows us to identify new effects related to the atomic structure of graphene, strengths of the electric field, valley mixing, and valley asymmetry. 

\end{abstract}



\maketitle

\section{Introduction and model description}

There has been a series of theoretical works over the last few years devoted to the study of the energy structure of electric field-defined topological quantum dots in bilayer graphene (BLG QD) \cite{QD_PRB_2021,QD_NJP_2022,QD_PSS_2022}. 
On the one hand, they respond to the significant progress that has been made recently in the experimental realization and manipulation of various BLG QDs  \cite{Eich_PRX_2018,Kurzman_PRL_2019,Banszerus_NL_2020,Ge_NL_2020,Banszerus_PRB_2021,Eich_NL_2018,Banszerus_NL_2020_2,Clerico_SR_2019}. 
On the other hand, they are part of the growing interest 
in graphene-based zero-dimensional structures due to their potential applications in e.g., quantum information processing \cite{Peeters_NL_2007,Chico_PRB_2010,Peeters_PRB_2015,Recher_PRB_2009,Recher_Nanotechnology_2010,Allen_NC_2012,Potasz_NL_2015,Peeters_PRB_2016,Kurzman_NL_2019,Wang_Nanotenology_2021,Garreis_NPhys_2024,Sadecka_NL_2023,Korkusinski_NL_2023}. 

A perpendicular electric field applied to the Bernal-stacked bilayer graphene opens a tunable energy gap. A spatially varying field can confine electrons in finite regions of the BLG. The most intriguing configuration is when the field changes only the sign in different regions. In such a case, a domain wall is formed between the regions, at which one-dimensional (1D) chiral gapless modes are localized \cite{Martin_PRL_2008}. When the region where the field changes its polarization is finite, the 1D modes discretize to a series of bound states localized at the edge of the region. They are called topologically confined states. 
The cited works \cite{QD_PRB_2021,QD_NJP_2022,QD_PSS_2022}, although using only simple one-valley continuum approximations, are the first 
to calculate these states for various electrically defined quantum dots in the BLG. 
 
\begin{figure}[ht]
\centering
\includegraphics[width=\columnwidth]{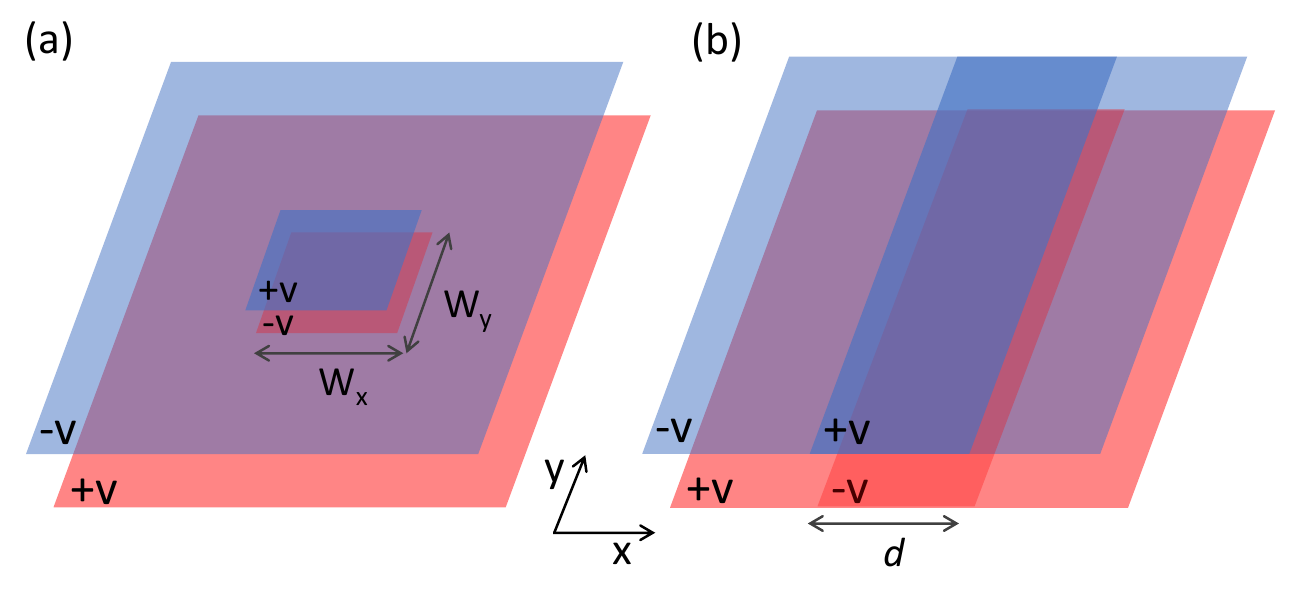}
\caption{\label{fig:first}
Schematic representations of electrically defined rectangle BLG QD (a) and a system of two parallel EFWs 
separated by $d$ (b). 
The dot is defined as a region in gated BLG, where the voltages $\pm V$ 
applied to the layers (red and blue) are reversed. 
}
 \end{figure}

However, the continuum models do not take into account the atomic and periodic structure of the BLG, do not recognize different directions in graphene, are mostly limited to weak fields, and cannot account for possible valley interaction. 
This paper aims to go a step forward to fill this gap and calculate the energy structure of electrically defined rectangular BLG QDs, as schematically visualized in Fig. \ref{fig:first} (a), in an atomistic tight-binding approach (TB). 
To investigate the influence of the BLG atomic structure on the edge-confined states in electrically defined QDs, we perform calculations for rectangular dots oriented in the armchair and zigzag directions. 
This is, however, challenging, since electrically defined BLG QDs of $\sim$50 nanometers in size would require performing calculations on systems comprising almost half a million atoms. 

To address this challenge, we simplify the model and begin by examining two parallel electric-field domain walls (EFWs) created by reversing the electric field twice, as shown schematically in Fig. \ref{fig:first} (b).
First, infinite-length domain walls with one-dimensional topological bands localized at the walls are considered. Then, walls of finite length are taken into account, which lead to band discretization and the formation of topologically confined bound states. 
The parallel walls act as opposite sides of the rectangular QD. 
The 1D topological bands localized at perpendicular sides of the QD discretize independently, which simplifies the calculations. 
The TB calculations reveal several new effects that are not accessible in simpler models. 
We find that the bound states localized at the zigzag edges are affected by the cone asymmetry. This makes them behave differently from the states localized at the armchair edges, as the dot size increases. 
We also show that on the zigzag edges, bound states with energies independent of the QD size can additionally appear.

\section{Computational method}
The Hamiltonian in the $\pi$-electron tight-binding approximation that we use in calculations is 
$$ 
H=t_{i/e}{\sum_{\left\langle i,j \right\rangle }{c_i^{\dagger}c_j }} + {\rm{H.c}}
$$
where $c_i^{\dagger},\left(c_i\right)$ are the creation and (annihilation) operators at the site $i$ and the symbol 
$\left\langle i,j \right\rangle $ restricts summation to the nearest neighbors. We use standard intra-layer and inter-layer hopping parameters $t_i = 2.7\,$eV and $t_e=0.27\,$eV, respectively \cite{WJ_2D_Materials_2018,WJ_Materials_2025}. 
The boundary conditions (BC) applied in the calculations for different structures are discussed in the sections where the results are presented. 

The perpendicular electric field is considered by adding gate voltages $\pm V$ to the upper and lower BLG layers. The calculations are performed for two different gates, $V=\pm 0.1\,$V and $V=\pm 0.5\,$V. In the first case, the voltage applied between the layers is smaller than the interlayer interaction energy $t_e$, while in the second case it is larger. 
This choice is dictated by previously observed gate control of the layer localization of gapless states at domain walls formed by the stacking order change in multilayer graphene \cite{WJ_2D_Materials_2018,WJ_TLG_PRB_2020}. It may influence band crossing in the case of two parallel domain walls. 
Since the aim of this work is to show that the atomistic approach yields some new results compared to continuum one-valley approximations, the domain walls are, for simplicity, formed by a simple sign reversal of the voltage applied.

\section{Results and discussion}
\subsection{Two parallel EFWs}

First, we investigate the system of two parallel EFWs, as schematically shown in Fig. \ref{fig:first} (b). The domain walls can extend along the armchair or zigzag direction and are separated by a distance $d$ measured in the number of unit cells in a given direction. 
The unit cells contain eight carbon atoms (four in each layer). 
The length of unit cells in the zigzag and armchair directions is $a_{z}=\sqrt{3}a_{C-C}$ and $a_{a}=3a_{C-C}$, respectively, where $a_{C-C}$ is the distance between two carbon atoms in a layer. 
To calculate the energy bands of such systems, we must impose boundary conditions away from the domain walls. 
The BCs can be open (the edges in $x$-direction of the entire system shown in Fig. \ref{fig:first} (b) not connected) or closed (the edges connected). 
We choose the size of the system in the direction perpendicular to EFWs, large enough to ensure that the results are practically independent of the size and BC.
In the case of EFWs in the zigzag direction, we use closed BC to get rid of zigzag edge states.

\begin{figure}[ht]
\centering
\includegraphics[width=\columnwidth]{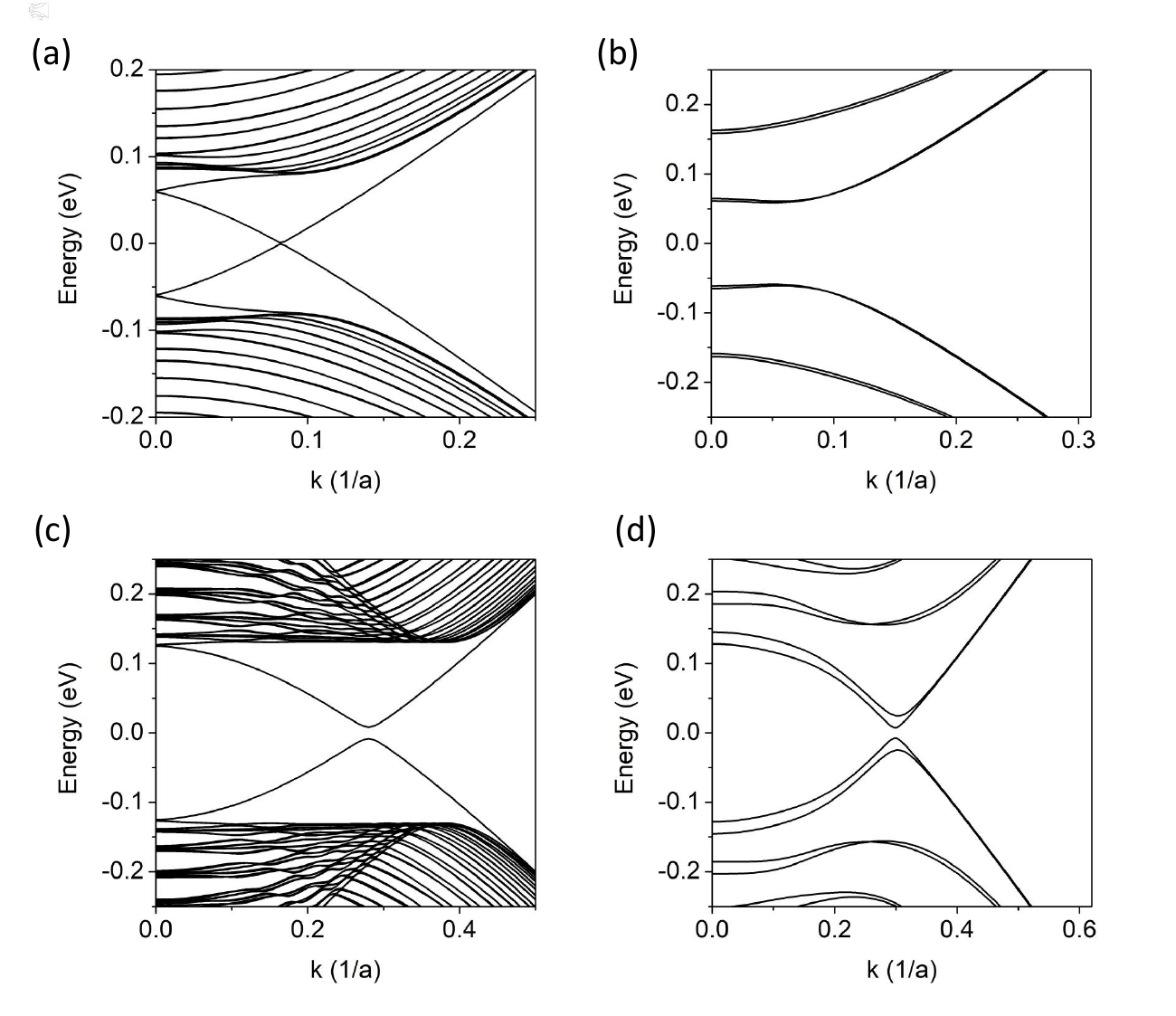}
\caption{\label{fig:second}
Energy bands close to $\Gamma$ ($k=0$) for two parallel EFWs along the armchair direction. In (a) and (c), the walls are separated by 140 unit cells; in (b) and (d), they are separated by 20 unit cells. Voltage applied to the layers is $V=\pm0.1\,$V in (a) and (b), and $V=\pm0.5\,$V in (c) and (d). 
}
 \end{figure}

Each domain wall in a system of two parallel EFWs introduces in each valley (Dirac cone) a pair of topological one-dimensional gapless states of a given slope (momentum). 
If the EFW lies along the armchair direction, the valleys $K$ and $K^{\prime}$ overlap at the $\Gamma$ point, so even for a single EFW, there are two pairs of gapless bands, each with opposite slope. 
Thus, when two EFWs are very well separated, all these bands are two-fold degenerate, and the crossing point is four-fold degenerate. This is seen in Fig. \ref{fig:second} (a) for $ d=140\,a_z$ and small voltage $V=0.1\,$V. 
Higher voltage, $V=0.5\,$V, acts as a strong perturbation and removes the degeneracy of the pair of bands at the same domain wall, i.e., corresponding to $K$ and $K^{\prime}$ (see panel (c) at $k\approx 0.3$), but the bands maintain two-fold degeneracy due to the large separation distance between the domain walls. When the separation distance is small, $d=20\,a_z$ (Fig. \ref{fig:first} (b) and (d)), the topological bands slightly split and anti-cross due to the interaction of bands localized at different domain walls.

\begin{figure}[ht]
\centering
\includegraphics[width=\columnwidth]{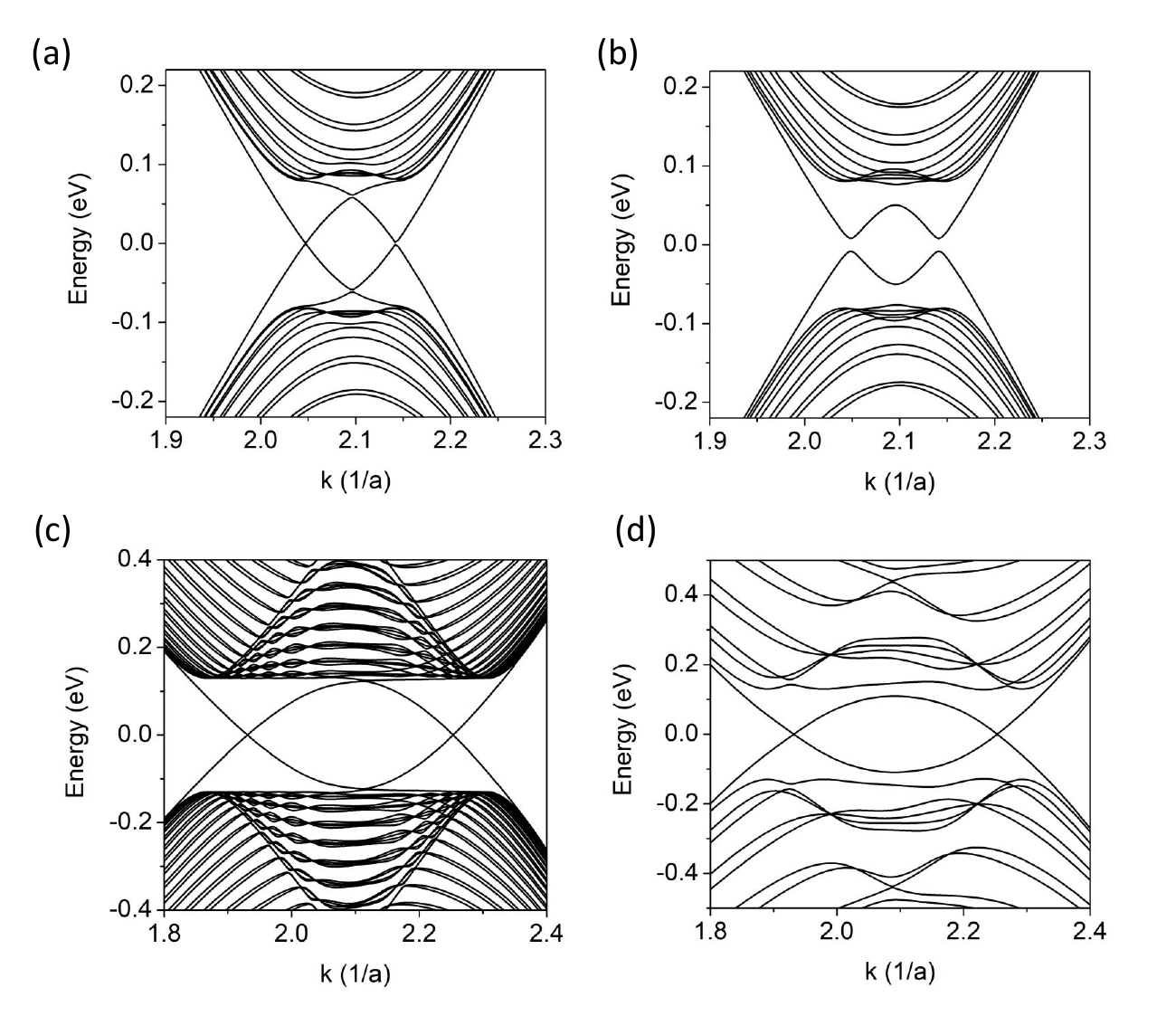}
\caption{\label{fig:third}
Energy bands close to $k=\frac{2\pi}{3a_z}$ for two parallel EFWs along the zigzag direction. In (a) and (c), the walls are separated by 80 unit cells; in (b) and (d), they are separated by 20 unit cells. Voltages applied to the layers are $V=\pm0.1\,$V in (a) and (b), and $V=\pm0.5\,$V in (c) and (d). 
}
 \end{figure}

The energy bands of two EFWs along the zigzag direction, separated by $d=80$ and $d=40$ unit cells for two gates $V=0.1\,$V and $V=0.5\,$V, are shown in Fig. \ref{fig:third}. 
In this case, all the 1D gapless bands are non-degenerate, since they correspond to a single $K$; the pairs of bands with opposite slopes originate from different domain walls. 
For $V=0.1\,$V and a smaller separation distance between the EFWs ($d=20\,a_a$, panel b), the topological bands slightly anti-cross. 
However, for $V=0.5\,$V the bands cross even for small $d=20\, a_a$. 
This is because at high voltages, the crossing bands localize mainly in different layers and non-connected nodes of the layers \cite{WJ_2D_Materials_2018}. 

\subsection{Two parallel EFWs of finite width} 

In this section, we investigate double EFW of finite length $W$, which we call {\it{the width}} to be consistent with the description of finite-size QD (see Fig. \ref{fig:first} a). To see how the topological bands are discretized for a domain wall of finite width, we select the discrete values of the wave vector $k$ in the calculations performed in the previous section, i.e.,  
\begin{equation}	 
k_n = \frac{n \pi}{W}, \;  n=1,2,..,N
\end{equation}
where $W=Na_{a/z}$ and $a_{a/z}$ equals $a_a$ or $a_z$ for EFW in the armchair or zigzag direction, respectively. 
This is equivalent to closed BC applied at $W$ \cite{Pelc_PRB_2015}.
Those $E_n$ that lie within the energy gap correspond to discrete energy levels arising from topological gapless modes. The corresponding wave functions are localized at the EFWs.

\begin{figure}[ht]
\centering
\includegraphics[width=\columnwidth]{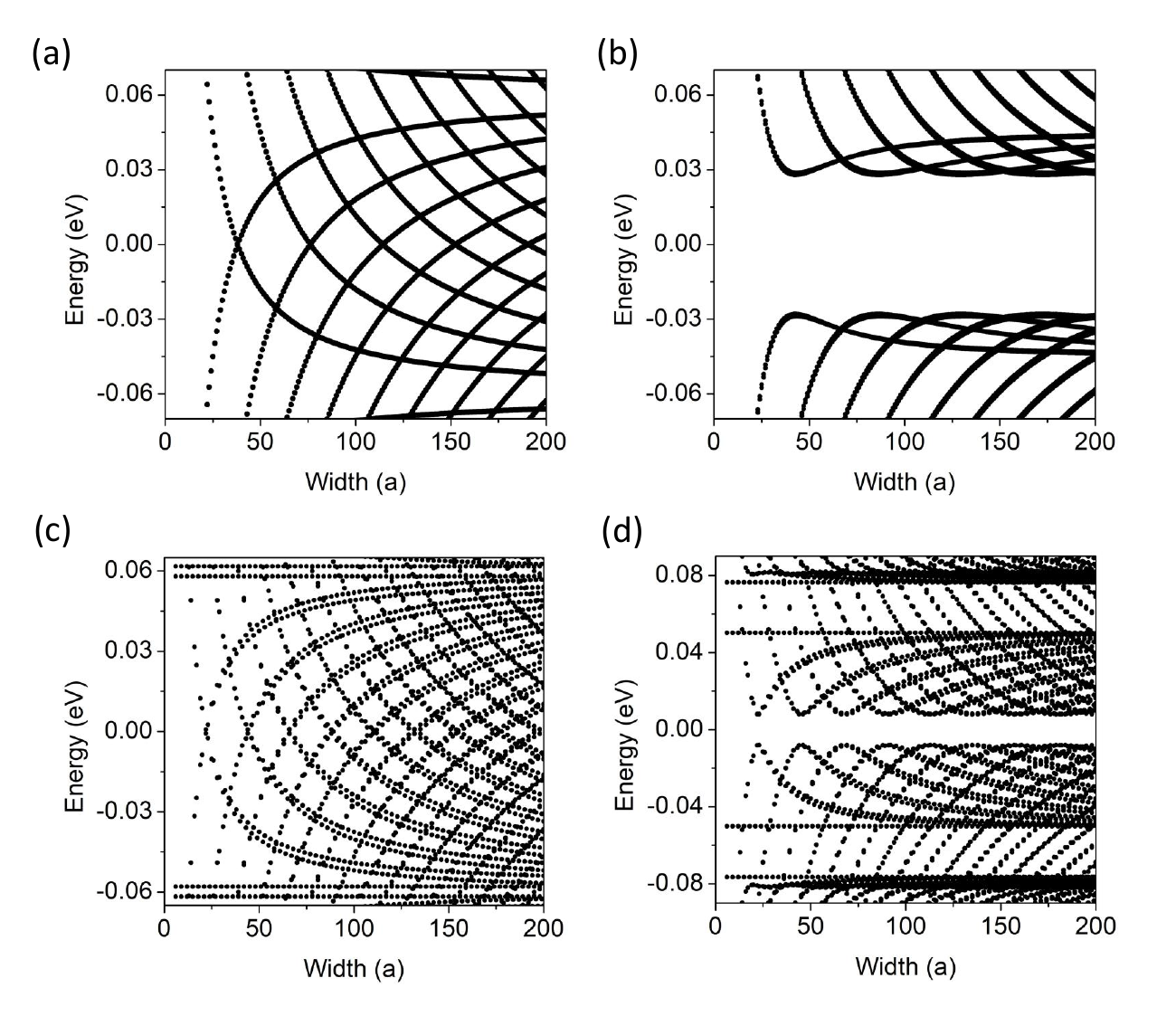}
\caption{\label{fig:fourth}
Energy levels of two parallel EFWs of finite width $W$. The voltages applied to the layers are $V=\pm0.1\,$V. Upper panels: EFWs along the armchair direction, separated by 160 unit cells (a) and by 40 unit cells (b). Lower panels: EFWs along the zigzag direction, separated by 80 and 40 unit cells in (c) and (d), respectively. 
}
 \end{figure}

The discrete energy levels of two EFWs along the armchair direction, as a function of the domain wall width $W$, for $V=0.1\,$V and two separations between the walls $d=160\,a_z$ and $d=40\,a_z$ are shown in Fig. \ref{fig:fourth} (a) and (b), respectively. 
For each $d$, the energy levels arrange in a kind of {\it{branches}} for increasing $W$ \cite{QD_NJP_2022,QD_PSS_2022}. 
The branches cross and have mirror symmetry with respect to $E=0$, the same as 1D gapless bands in Fig. \ref{fig:second}. 
For a small separation between the domain walls, $d=40\,a_z$, the absolute energy gap appears. 
For very large widths $W$, the branches will converge to the energies corresponding to $k=0$ of the case of infinitely long domain walls (i.e.  $E\approx \pm 0.6\,$eV in Fig. \ref{fig:second} (a) for $d=140\,a_z$). 
It is also worth noting that all the discrete energy levels are doubly degenerate, as are their corresponding 1D topological bands. 

The discrete energy levels of two finite EFWs along the zigzag direction, as a function of the domain wall width $W$, for $V=0.1\,$V and two separations between the walls $d=80\,a_a$ and $d=40\,a_a$ are shown in Fig. \ref{fig:fourth} (c) and (d), respectively. 
Although the general pattern of the energy branches is similar to that of two EFWs along the armchair direction, there are important differences that need to be analyzed and explained. 
These main differences are: (i) the appearance of {\it{flat}} branches, and (ii) the duplication of branches that looks like splitting. 

As for the flat branches, they result from the position of the Dirac cone, which, in the zigzag direction, is at $k=\frac{2\pi}{3a_z}$. Thus, the flat branches appear whenever the width $W=Na_z$ is a multiple of 3 in equation (1), i.e., $N=3M$, where $M$ is an integer. 
In such a case, the cut of topological bands always yields the same energy for all $n=\frac{2}{3}N$. This does not happen when EFW is in the armchair direction, since even for a large $W$, we never reach the cone center at $k=0$. 

As for branch duplication, it is not a true splitting, since the topological bands for EFW in the zigzag direction are not degenerate. This effect results from cone asymmetry. 
Indeed, in the armchair direction, the cone at $\Gamma$ is symmetric with respect to $k=0$, while in the zigzag direction, the cone at $k_0=\frac{2\pi}{3a_z}$ is not symmetric with respect to $k_0$. 
This asymmetry is an intrinsic property of the valley when the graphene $k$-space is projected along the line corresponding to the zigzag direction. 
The slope of $E(k_0+\delta k)$ is different 
than the slope $E(k_0 -\delta k)$.
It is like moving on the graphene 2D energy surface from $K$ to $M$ on one side of $k_0$ and from $K$ to $\Gamma$ on the other.
This cone asymmetry is well seen in Fig. \ref{fig:third} (c) and (d)
\cite{notka_asym}. 
The asymmetry of the band continua translates into asymmetry of the topological bands. In effect, these bands are slightly shifted and do not cross exactly at $k=\frac{2\pi}{3a_z}$. 
As a consequence, for the growing width $W$, the discrete energy levels $E_n$ do not arrange monotonically in the branches. This is clearly seen in the case of flat branches. 
If the cone were exactly at $k_0=\frac{2\pi}{3a_z}$, the flat branches wouldn't split. But since the crossing point is slightly shifted from  $k_0$, the pair of close energy levels appears when $k_n=\frac{2\pi}{3a_z}$. 

It is worth emphasizing that none of these effects could be visible in continuum single-valley approximations, because they essentially reflect only the armchair direction with a symmetric cone.

\begin{figure}[ht]
\centering
\includegraphics[width=\columnwidth]{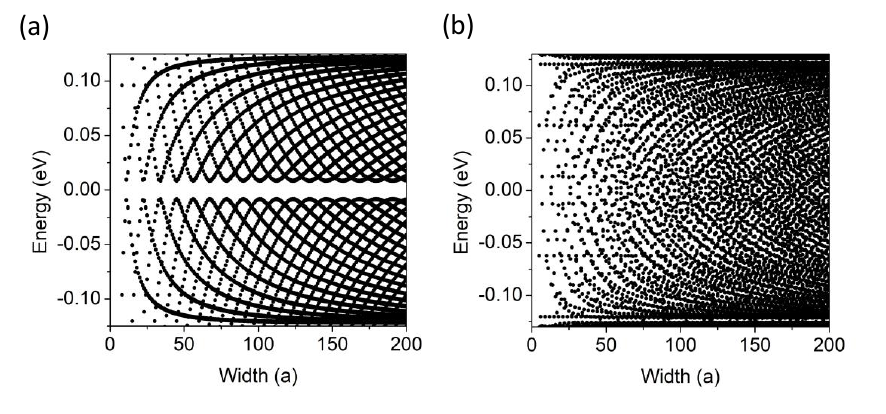}
\caption{\label{fig:fifth}
Energy levels of two parallel EFWs of finite width $W$. The voltages applied to the layers are $V=\pm0.5\,$V. (a) EFWs in armchair direction separated by 80 unit cells; (b) EFWs in zigzag direction separated by 40 unit cells 
}
 \end{figure}

In Fig. \ref{fig:fifth}, branches of discrete energy levels for $V=0.5\:$V are presented. Panel (a) shows energy levels of two EFWs of finite width $W$ in the armchair direction, separated by $d=80$\:$a_z$.
The absolute energy gap arises from partial splitting of the crossed 1D bands, as described in the previous section and shown in Fig. \ref{fig:second} (c) and (d).
Panel (b) shows energy levels of two EFWs of finite width $W$ in the zigzag direction, separated by $d=40\,a_a$. Branch duplication is clearly seen, and flat branches at $E\approx \pm 0.14\,$V are also visible (the second pair of this branch doublet is already embedded in the continuum, see Fig. \ref{fig:third} d). 
Other flat branches, closer to $E=0$ and less dense, are also seen. 
They appear when $W=Na_z$ is a multiple of another integer, $N=mM$, and $k_n=\frac{n\pi}{mMa_z}$ crosses the 1D topological modes in the energy gap. For example, $m=5$ gives $k_n\approx 1.9$ in the units of $1/a_z$ (see Fig. \ref{fig:third} c and d) and the same energies close to $E\approx \pm 0.6\,$eV for all $n=\frac{3}{5}N$. 

\subsection{Rectangual QD}

\begin{figure}[ht]
\centering
\includegraphics[width=\columnwidth]{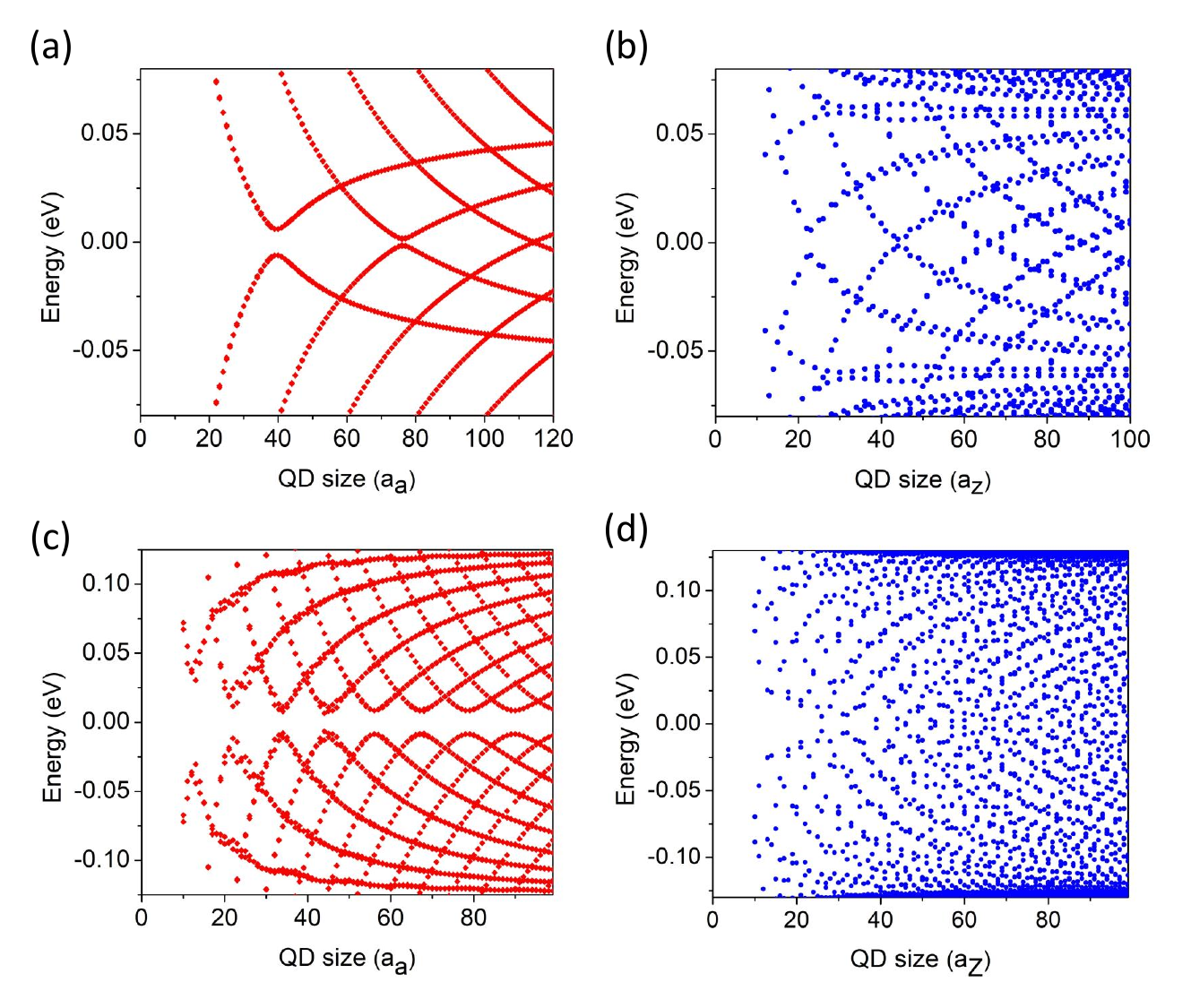}
\caption{\label{fig:sixt}
Energy levels of rectangular QDs with edges oriented along the armchair and zigzag directions. Left and right panels show energy levels originating from the armchair and zigzag edges, respectively. Voltages applied to the layers are $V=\pm0.1$ in (a) and (b), and $V=\pm0.5$ V in (c) and (d).  
}
 \end{figure}

The TB calculation for electrically-defined QDs would require considering a system much larger than the QD itself, i.e., a rectangle embedded in BLG with reversed gates, as shown in Fig. \ref{fig:first}(a). 
For a quantum dot of the order of a hundred units in each direction, the entire system would contain hundreds of thousands of atoms. 
We avoid large-scale calculations by building on the findings of the previous sections. The $k$-vectors of 1D bands localized at perpendicular edges of QD are also perpendicular and have different values. 
Thus, without much loss of generality, we can assume that the discretization of 1D topological gap modes localized at the armchair- and zigzag-oriented sides of the rectangle is done independently. 
This is like wavefunction factorization used in continuous models \cite{QD_PRB_2021}.

Therefore, by selecting discrete values of $k$ defined in equation (1) for $W=W_a$ in the case of two EFW in the armchair direction separated by $d=W_z$, we get discrete energy levels originating from 1D bands localized at the armchair sides of the rectangular QD $W_a \times W_z$. 
Similarly, selecting discrete values of $k$ for $W=W_z$ in the case of two EFW in the zigzag direction separated by $d=W_a$, we get discrete energy levels originating from 1D bands localized at the zigzag sides of the QD.
This is exactly what we have done in the previous section.
Figures \ref{fig:fourth} and \ref{fig:fifth} show the energy levels of topological bound states on one side of the quantum dot (as a function of the width of that side), while keeping the width $d$ of the other side constant.

Now, we consider a rectangular QD whose sides change simultaneously with the change in the dot size. 
Their widths, measured in units of $a_a$ and $a_z$, are $W_a$ and $W_z$, respectively.   
For simplicity, we consider $W_a=W_z$, but since the units $a_a$ and $a_z$ are different, the dot is rectangular in nm.

In Fig. \ref{fig:sixt} we present discrete energy levels of rectangular QD of the size $W_a \times W_z$ for two different voltages, $V=0.1\,$V in panels (a) and (b), and $V=0.5\,$V in panels (c) and (d). The energies corresponding to states localized at the armchair edges are shown in panels (a) and (c); those localized at the zigzag edges are presented in panels (b) and (d) \cite{notka_separate_panels}. These spectra are similar to those presented in Figs. \ref{fig:fourth} and \ref{fig:fifth} for two EFW of finite width. 
In particular, the absolute energy gap in branches originating from the armchair edge, as well as doubled branches of bound states localized at zigzag edges, is seen. 
Furthermore, some traces of flat branches are visible in panel (b), although they differ slightly from those in Figs. \ref{fig:fourth} and \ref{fig:fifth} because the separation distance $d$ between the EFWs now changes simultaneously with the width $W$.

\section{Summary and Conclusions}

We have studied electric field-defined rectangular quantum dots in bilayer graphene with the dot sides oriented along the armchair and zigzag directions. 
The calculations have been performed in the tight-binding approach.
This has allowed us to reveal new features of the topological spectra of edge-confined states in electrically defined quantum dots in BLG, compared to simple one-valley continuum approximations.

We have begun analysis by considering two parallel domain walls defined by a change of the field sign.
Electric-field domain walls introduce and localize a pair of gapless 1D chiral modes, one pair in each valley $K$ and $K^{\prime}$, with opposite valley momenta. 
For domain walls along the armchair and zigzag directions, the valleys appear at $\Gamma$ and $\pm \frac{2\pi}{3a}$, respectively; they also differ in the cone symmetry and degeneracy of the 1D modes.
This has a fundamental impact on the spectra of topologically confined states, which arise from the discretization of chiral modes for finite-width domain walls, i.e., the edges of quantum dots. 

The energies of topologically confined states arrange themselves into branches as the size of the quantum dot increases. 
We have found that the branches behave differently for bound states localized at the zigzag and armchair edges. 
The branches of states originating from the zigzag edge are doubled, being a direct consequence of the cone asymmetry at $\frac{2\pi}{3a_z}$. 
Moreover, when the width $W$ of the zigzag edge increases, some energy levels repeat frequently, forming a kind of flat branches. It happens, for example, when $W$ is a multiple of 3. 
It is worth noticing that the edge of QD of any shape will contain zigzag components, and therefore, its spectrum of topologically confined states should also exhibit these features.


\begin{thebibliography}{99}
\bibitem{notka_asym} The cone asymmetry is more pronounced for higher gates, because of the larger overlap of the valence and conduction band continua.
\bibitem{notka_separate_panels} The energy levels originating from different QD edges are presented separately to show their differences.
\end{thebibliography}
\end{document}